# Assessing Co-Authored Papers in Tenure Decisions: Implications for Research Independence and Career Strategies in Economics

Lekang REN

The IPE Thrust, Society Hub, HKUST(GZ)

Danyang XIE

The IPE Thrust, Society Hub, HKUST(GZ)

January 2025

**Abstract**

In tenure decisions, the treatment of co-authored papers often raises questions about a candidate's research independence. This study examines the effects of solo versus collaborative authorship in high-profile Economics journals on long-term academic success. Our findings confirms the traditional belief that solo-authored publications significantly enhance long-term research output and citation impact compared to collaborative efforts. However, relative to solo-authored papers, international collaborations have a less negative impact on long-term success than national and institutional collaborations. Temporal trends highlight the increasing importance of diverse and international collaborations. These insights provide actionable guidance for tenure committees on evaluating co-authored work and for researchers on optimizing their publication strategies.

**Introduction**

The competitive nature of academia places early-career researchers under immense pressure to publish in prestigious journals to secure tenure and long-term academic success (Hou et al 2022). Lee (2019) analyzed the total citations received during the second four-year phase of researchers' careers to determine how early career impact influences later success. A critical aspect of tenure decisions is the evaluation of co-authored papers, which raises important questions about a candidate's research independence. This study investigates the long-term academic outcomes of solo versus collaborative authorship among economists and provides insights into how tenure committees should assess co-authored work. Additionally, it offers strategic guidance for researchers on optimizing their publication strategies based on these assessment criteria.

Our analysis confirms the traditional belief that solo-authored papers significantly enhance long-term productivity and citation impact compared to collaborative efforts. Economists who published solo early in their careers demonstrated higher average citation counts and paper counts over time, highlighting the importance of independent research contributions (Horta and Santos, 2010). However, our findings also show that not all collaborations are equal. Relative to solo-authored papers, international collaborations have a less negative impact on long-term success than national and institutional collaborations. This suggests that while solo-authored publications remain crucial, international collaborations should be valued more favorably in tenure assessments compared to other forms of collaboration. In our discussions, we do not explore whether solo or co-authored papers are of higher quality. This question is intriguing in its own right. According to Bridgstock (1991), the answer remains unresolved and may vary by field, though his findings suggest that solo-authored papers often come out on top. In principle, we can examine whether a finer classification of collaborative work into international, national, and institutional collaborations sheds new light on this question.

Temporal trends further underscore the increasing importance of diverse and international collaborations. Post-2000, the penalties for institutional collaborations have become more pronounced, reflecting the evolving landscape of academic research where interdisciplinary and cross-border collaborations are increasingly valued.

For tenure committees, these insights suggest that greater weight should be assigned to solo-authored papers to recognize research independence while also appreciating the value of international collaborations. For researchers, prioritizing solo-authored publications and engaging in strategic international collaborations can optimize long-term academic success and meet tenure expectations. These findings provide actionable guidance for both tenure committees and researchers in navigating the complexities of academic publishing and career advancement.

**Methods**

Using a comprehensive dataset, we analyzed the publication records of 6,360 economists who published in 68 leading economics journals between 1980 and 2014. Our primary objective was to assess the impact of early solo and collaborative publications on long-term academic productivity and citation impact. To achieve this, we ran regression models that allowed us to examine the relationship between authorship type and long-term outcomes while controlling for various confounding factors.

Each economist's publication history was meticulously recorded, noting whether their early works were solo-authored or collaborative. We categorized collaborations into three types: institutional (co-authors from the same institution), national (co-authors from different institutions within the same country), and international (co-authors from different countries). This categorization enabled us to investigate the differential impacts of various types of collaborations.

To ensure the robustness of our findings, we controlled for several key variables. Demographic controls include **Gender**, represented as a dummy variable. Educational background is reflected through the **Institution Score**, which is derived from the annual QS Score of the author's affiliated institution. This score, ranging from 20 to 100 (with scores below 20 recorded as 0), encompasses factors such as academic and employer reputation, student-to- faculty ratio, and citations per faculty (QS Quacquarelli Symonds Limited, 2023). Academic performance metrics encompass several variables: **Journal Citation Score**, sourced from the Scopus database, which provides a continuous measure of journal quality and output; and **Duration**, which measures the span between the author's first and most recent publication, thus reflecting the length of dedicated academic engagement in economics.

Our regression models included both linear and logarithmic transformations of citation counts and publication numbers to address potential skewness in the data. This methodological approach allowed us to draw nuanced conclusions about the relative impacts of solo and collaborative authorship on the long-term academic trajectories of economists.

**Results**

Results are shown in Table 1. First, concerning Solo vs. Collaborative Authorship. Solo-authored papers significantly enhance long-term productivity and citation impact, outperforming collaborative efforts. Economists with early solo-authored publications show higher average citation counts and paper counts over time.

Second, when focusing on collaborations, we find that relative to solo-authored papers, international collaborations have a less negative impact on long-term success than national and institutional collaborations.

Temporal Trends: The negative effects of homogeneous collaborations have deepened over time, highlighting the growing importance of diverse and international partnerships. Post-2000, the penalties for institutional collaborations have increased.

**Concluding Remarks**

For Tenure Committees: The findings suggest that solo-authored publications are strong indicators of a candidate's research independence and long-term academic potential. Tenure committees should recognize the significant contributions of solo-authored papers and consider them as a key factor in tenure decisions. However, collaborative work, particularly international collaborations, should be less heavily discounted compared to national and institutional collaborations due to their relatively higher positive impact. Institutional collaborations should be the most-heavily discounted.

For Individual Researchers: If institutions follow the recommendations above in their assessment approach, early-career researchers should: Prioritize solo-authored publications to establish a strong academic identity and demonstrate research independence. Engage in strategic collaborations, particularly international ones, to enhance their visibility and influence without overshadowing their individual contributions. Balance solo and collaborative efforts to optimize their long-term academic success and meet tenure expectations.

**Table 1**: Regression Result for Citation Count and Paper Count

| VARIABLES | Log Citation Count | | | Log Paper Count | | |
|---|---|---|---|---|---|---|
| | Model 1 | Model 2 | Model 3 | Model 4 | Model 5 | Model 6 |
| Institutional Collaboration | -0.550 *** | -0.358 *** | -0.343 *** | -0.181 *** | -0.082 *** | -0.082 *** |
| | (0.069) | (0.057) | (0.058) | (0.025) | (0.022) | (0.022) |
| National Collaboration | -0.229 *** | -0.138 *** | -0.218 *** | -0.124 *** | -0.049 *** | -0.064 *** |
| | (0.039) | (0.033) | (0.032) | (0.017) | (0.014) | (0.015) |
| International Collaboration | -0.451 *** | -0.224 *** | -0.142 *** | -0.128 *** | -0.001 | 0.048 |
| | (0.037) | (0.031) | (0.032) | (0.016) | (0.013) | (0.014) |
| Gender (Male) | | 0.0464 | 0.072 *** | | 0.1001 *** | 0.103 *** |
| | | (0.033) | (0.032) | | (0.013) | (0.013) |
| Duration | | 0.064 *** | 0.062 *** | | 0.033 *** | 0.033 *** |
| | | (0.002) | (0.003) | | (0.001) | (0.001) |
| Journal Score | | 0.062 *** | 0.005 *** | | 0.002 ** | 0.001 |
| | | (0.003) | (0.003) | | (0.001) | (0.001) |
| Institution Score | | 0.009 *** | 0.010 *** | | 0.005 *** | 0.003 *** |
| Constant | 5.245 *** | 3.05 *** | 3.245 *** | 2.485 *** | 1.548 *** | 1.561 *** |
| | (0.023) | (0.051) | (5.108) | (0.010) | (0.021) | (0.107) |
| R-squared | 0.028 | 0.339 | 0.379 | 0.017 | 0.304 | 0.323 |
| Country FE | NO | NO | YES | NO | NO | YES |
| N | 6360 | 6360 | 6360 | 6360 | 6360 | 6360 |

Note: Standard errors in parentheses, Model (1)(4): Only consider Author Type , Model (2)(5): Consider controls without Country FE, Model (3)(6): Consider controls without Country FE. Robust standard errors in parentheses *** p<0.01, ** p<0.05, * p<0.1.

# Cited References in this Paper

Note: For the sake of brevity of this paper and keep to the point, we shortened the introduction and literature review substantially and only four most relevant papers are cited explicitly in this paper. Complete references cited in the original work are also provided below for those who may be interested.

# Complete References